\documentclass[journal]{IEEEtran}
\usepackage{lineno,hyperref}
\usepackage[dvipsnames]{xcolor}
\usepackage[ruled,vlined]{algorithm2e}
\usepackage{amsmath,amssymb,amsthm} 
\usepackage{graphicx}
\usepackage{enumitem}
\usepackage{float}
\usepackage{caption}
\usepackage{subcaption}
\usepackage[normalem]{ulem}
\usepackage{cite}
\usepackage{changes}
\usepackage{yhmath}
\usepackage{fancyhdr}

\newtheorem{theorem}{Theorem}
\newtheorem{lemma}{Lemma}

\newtheorem{corollary}{Corollary}
\newtheorem{definition}{Definition}
\newtheorem{remark}{Remark}

\newtheorem*{notation*}{Notation}
\DeclareMathOperator*{\argmin}{arg\,min}

\DeclareMathOperator*{\eig}{eig}
\DeclareMathOperator{\E}{\mathbb{E}}

\newcommand{\RMA}[1]{{\color{black}#1}}

\newcommand{\LAI}[1]{{\color{black}#1}}
\newcommand{\JB}[1]{{\color{black}#1}}

\fancypagestyle{firstpage}
{
    \fancyhead[L]{This work has been submitted to Computers \& Chemical Engineering for possible publication.}    
    \fancyhead[R]{}
}
\begin{document}
\title{Discrete-time Contraction-based Control of Nonlinear Systems with Parametric Uncertainties using Neural Networks}
\author{Lai Wei,
        Ryan McCloy,
        and Jie Bao
\thanks{Lai Wei, Ryan McCloy and Jie Bao are with the School of Chemical Engineering, The University of New South Wales (UNSW Sydney), Sydney, NSW 2052, Australia. E-mail: lai.wei1@unsw.edu.au, r.mccloy@unsw.edu.au and j.bao@unsw.edu.au (corresponding author).}
\thanks{This project is partially supported by Australian Research Council (ARC) Discovery Projects DP180101717 and ARC Research Hub IH180100020.}}
\maketitle

\begin{abstract}
\RMA{In response to the continuously changing feedstock supply and market demand for products with different specifications, the processes need to be operated at time-varying operating conditions and targets (e.g., setpoints) to improve the process economy, in contrast to traditional process operations around predetermined equilibriums. In this paper, a contraction theory-based control approach using neural networks is developed for nonlinear chemical processes to achieve time-varying reference tracking. 
This approach leverages the universal approximation characteristics of neural networks with discrete-time contraction analysis and control. It  involves training a neural network to learn a contraction metric and differential feedback gain, that is embedded in a contraction-based controller. A second, separate neural network is also incorporated into the control-loop to perform online learning of uncertain system model parameters. The resulting control scheme is capable of achieving efficient offset-free tracking of time-varying references, with a full range of model uncertainty, without the need for controller structure redesign as the reference changes.} This is a robust approach that can deal with bounded parametric uncertainties in the process model, which are commonly encountered in industrial (chemical) processes. \JB{This approach also ensures the process stability during online simultaneous learning and control.} Simulation examples are provided to illustrate the above approach.


\end{abstract}
\IEEEpeerreviewmaketitle
\section{Introduction}
\thispagestyle{firstpage}

\IEEEPARstart{T}{he} process industry has seen increasing variability in market demand, with time varying specifications and quantity of products. As an example, there is a trend for produce-to-order operations in polymer \cite{zhang2016optimal}  and fine chemicals sectors. Variations are also evident in the specifications, costs and supply volume of “raw” materials and energy. Consequently, businesses in the process industry are required to have the flexibility to dynamically adjust the volume and specifications of products, and deal with diverse sources of raw materials to remain competitive \cite{chokshi2008drpc}. Traditionally, a chemical plant is designed and operated at a certain steady-state operating condition where the plant economy is optimized. The process industry is shifting from the traditional mass production to more agile, cost-effective and dynamic process operation closer to the market, which is the main driver for next-generation “smart plants”. As such, the control systems \LAI{for modern chemical processes} need to have the flexibility to drive a  process to any required time-varying operational target (setpoint) to meet the dynamic market demand, \LAI{with a stability guarantee, especially during the start-up or shut-down of processes.} 

As most chemical processes are inherently nonlinear, flexible process operation warrants nonlinear control as the target operating conditions may need to vary significantly to minimize the economic cost. \RMA{Control of nonlinear systems to track a time-varying reference profile and ensure stability (convergence) can be challenging \cite{manchester2017control}.
For general nonlinear systems (e.g., many of them appearing in chemical process systems), common nonlinear control approaches involve stabilization of nonlinear systems with respect to a given fixed equilibrium point, e.g., designing a control Lyapunov function for a given equilibrium point, and based on which, constructing a stabilizing control law. Therefore, every time the reference is changed, a new control Lyapunov function needs to be constructed and the control algorithm needs to be redesigned. This inherent lack of flexibility makes control Lyapunov function-based approaches infeasible for tracking arbitrary time-varying references (such as time-varying product specifications required by the market). To address the above challenge, contraction theory~\cite{lohmiller1998contraction,manchester2017control} was adopted to study the stability around (or contraction to) arbitrary references and design tracking controllers based on differential dynamics for time-varying (feasible) references of nonlinear systems with stability guarantees and without structural redesign (e.g.,~\cite{mccloybao2022, mccloy2021differential}).}

Neural networks have been used for nonlinear process modeling (e.g.,\cite{dai2013dynamic}) due to their ability to universally approximate arbitrary functions, adaptive control \cite{7078921,patino2000neural}, finding utilization in model-based control designs, such as model predictive control (MPC) (see, e.g., \cite{wang2015combined}). Neural networks have also been effectively used in learning control and adaptive control applications~\cite{ge2013stable}. A wide variety of neural network structures are available (e.g., Siamese networks \cite{sheng2019feature}), of which can greatly impact the  performance of the trained neural network. However, stability design for neural network-based control is still an open problem. In particular, there are very few results on setpoint-independent nonlinear control, especially when considering (embedding) the neural network within the closed-loop control system.
\RMA{This poses a significant obstacle for the neural network-based control approaches for chemical processes, which are mission critical. It is often impractical to operate a process in a large range of random operating conditions (with sufficient excitations) to produce process data to train neural networks to learn system dynamics (or even directly learn control laws). While such an exploration exercise is often performed for mechanical systems, it is generally infeasible for chemical processes, as many operating conditions may lead to poor product quality and/or inefficient material and energy usage with significant financial penalties. If process stability is not ensured, random process operating conditions can even cause severe safety risks (fires or explosions). Furthermore, chemical processes are typically designed based on known mechanisms and thus chemical process models are often available, although often with model uncertainties such as uncertain parameters within a known range (see, e.g.,~\cite{leitmann1993one}). As such, one promising approach in chemical process control~\cite{shin2019reinforcement} is to train neural network-based controllers using process data (input/output trajectories) generated from process models and further refine such neural networks using real-world process operating data. An important issue is ``safe exploration'' (during online simultaneous learning and control using neural networks), i.e., how can the controller explore system trajectories to obtain new operating data, such that the stimulus is  sufficient for online learning and refinement of such networks, yet simultaneously ensure process stability by exploiting known trajectories, during online simultaneous learning and control using neural networks. Inspired by \cite{shin2019reinforcement,vcrepinvsek2013exploration}, we aim to develop such a ``safe'' neural network-based online learning approach to determine uncertain system parameters by ``exploiting'' a neural network embedded controller that is designed offline via the contraction theory framework. 
}




To address the problem of controlling discrete-time nonlinear systems with parametric uncertainties, this article presents a systematic approach to obtaining a novel contraction-based controller with online parameter learning that embeds neural networks. 
Using properties of contracting discrete-time systems, we 
develop a novel approach that uses neural networks to synthesize DCCMs. This approach utilizes the nonlinear process model to train the DCCM neural network representations of the DCCM and differential feedback control gain (from which the control law can be computed). To train the DCCM neural network, a Siamese (or twin) neural network~\cite{sheng2019feature} is employed to ensure both steps in the state trajectory share the same state-dependent DCCM. A neural network-based learning module is also incorporated into the control-loop, such that any uncertain parameters can be identified online. Finally, conditions to ensure offset-free tracking (\LAI{after system parameters are correctly identified}), or bounded tracking (when parameters are modeled with known uncertainty bounds), to feasible references are derived. The resulting contraction-based controller embeds a neural network.


This article is structured as follows. Section \ref{sec:pro} formulates the problem of tracking-time varying references for uncertain nonlinear systems with neural networks in the closed loop, and the proposed approach which leverages the contraction theory framework. Section \ref{sec:DCCM} presents the main approach to designing a discrete-time contraction-based controller for discrete-time systems with parametric uncertainties using \LAI{DCCM neural networks and learning of uncertain parameters via estimation neural networks}. Section \ref{sec:design analysis} develops the conditions for the existence of contracting regions under the proposed neural network-based method of Section~\ref{sec:DCCM}. Section \ref{sec:sim} presents illustrative examples, followed by Section \ref{sec:conclusion} which concludes the article.

\begin{notation*}
    Function $f_k$ is defined as $f_k = f(x_k)$ for any function $f$, $\mathbb{Z}$ represents the set of all integers, $\mathbb{Z}^+$ represents the set of positive integers, $\mathbb{R}$ represents set of real numbers.
\end{notation*}
\RMA{
\section{Problem Formulation and Approach} \label{sec:pro}
\subsection{Contraction Theory Overview}\label{sec:discrete-time contraction analysis}

Contraction theory \cite{lohmiller1998contraction,manchester2017control} facilitates stability analysis and control of discrete-time nonlinear systems with respect to arbitrary, time-varying (feasible) references, without redesigning the control algorithm, through the study of corresponding displacement dynamics or \textit{differential dynamics}. The analysis and controller synthesis to ensure stability/contraction of the nonlinear system is simultaneously completed via discrete-time control contraction metrics (DCCMs). To introduce the contraction-based methodology, we firstly consider the discrete-time nonlinear control affine system without uncertainty (extended in later sections)
\begin{equation}\label{equ:pre cer sys}
    x_{k+1} = f(x_k) + g(x_k)u_k,
\end{equation}
where state and control are $x_k \in \mathcal{X} \subseteq \mathbb{R}^n$ and $u_k \in \mathcal{U} \subseteq \mathbb{R}^m$. The corresponding differential system of \eqref{equ:pre cer sys} is as follows
\begin{figure}
    \begin{center}
        \includegraphics[width=0.9\linewidth]{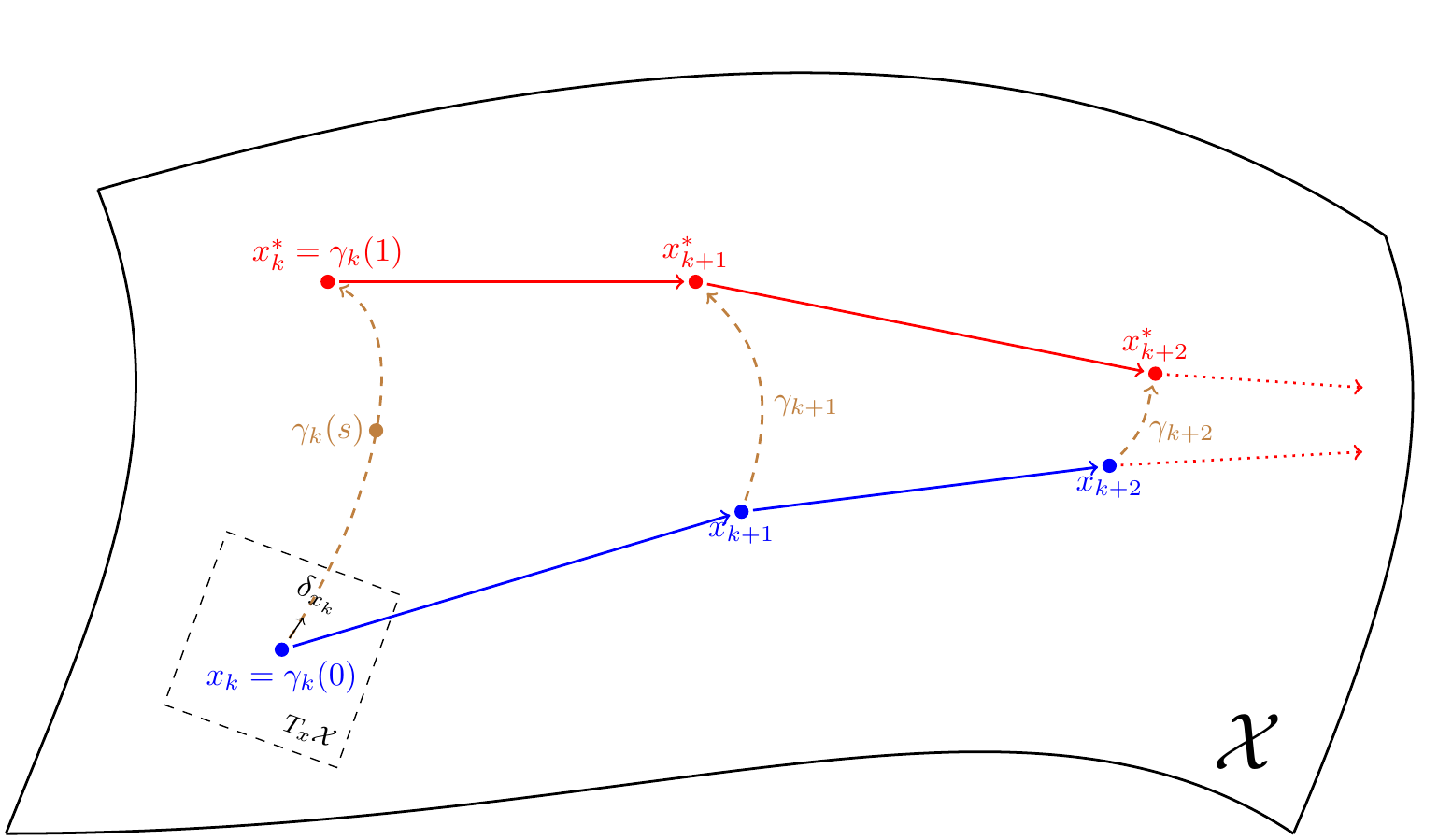}
        \caption{System trajectories along the state manifold $\mathcal{X}$.}
        \label{fig:rie_geo}
    \end{center}
\end{figure}
\begin{equation}\label{equ:differential dynamical system}
    \delta_{x_{k+1}} = A_k \delta_{x_k} + B_k \delta_{u_k},
\end{equation}
where Jacobian matrices of $f$ and $g$ in \eqref{equ:pre cer sys} are defined as $A_k = A(x_k):=\frac{\partial (f(x_k) + g(x_k)u_k)}{\partial x_k}$ and $B_k = B(x_k):=\frac{\partial (f(x_k) + g(x_k)u_k)}{\partial u_k}$  respectively, $\delta_{u_k} := \frac{\partial u_k}{\partial s}$ and $\delta_{x_k}:=\frac{\partial x_k}{\partial s}$ are vectors in the tangent space $T_x\mathcal{U}$ at $u_k$ and $T_x\mathcal{X}$ at $x_k$ respectively, where $s$ parameterises a path, $c(s): [0,1] \rightarrow \mathcal{X}$ between two points such that $c(0) = x, c(1) = x^*  \in \mathcal{X}$ (see Fig. \ref{fig:rie_geo}). Considering a state-feedback control law for the differential dynamics \eqref{equ:differential dynamical system}, 
\begin{equation}\label{equ:pre dif fed}
    \delta_{u_k} = K(x_k) \delta_{x_k},
\end{equation}
where $K$ is a state dependent function, we then have from \cite{lohmiller1998contraction,manchester2017control} the following definition for a contracting system. 

\begin{definition}\label{thm:pre ctr con}
    A discrete-time nonlinear system \eqref{equ:pre cer sys}, with differential dynamics \eqref{equ:differential dynamical system} and differential state-feedback controller \eqref{equ:pre dif fed}, is contracting, with respect to a uniformly bounded metric, $\alpha_1 I \leq M(x_k) \leq \alpha_2 I$ (for constants $\alpha_2 \geq \alpha_1 > 0$), provided, $\forall x \in \mathcal{X}$ and $\forall \delta_x \in T_x\mathcal{X}$, \begin{equation}\label{equ:pre ctr con} 
        (A_k+B_kK_k)^\top M_{k+1}(A_k+B_kK_k) - (1-\beta)M_{k} < 0,
    \end{equation}
    for some constant contraction rate $0 < \beta \leq 1$, where
    $M_k = M(x_k)$ and $M_{k+1} = M(x_{k+1})$. 
\end{definition}
A region of state space is called a contraction region if condition \eqref{equ:pre ctr con} holds for all points in that region. 
In Definition \ref{thm:pre ctr con}, $M$ is a metric used in describing the geometry of Riemannian space, which we briefly present here. We define the Riemannian distance, $d(x,x^*)$, as a measure for the tracking performance (convergence/stability of the state, $x$, to the reference, $x^*$), and is defined as (see, e.g., \cite{do1992riemannian})
\begin{equation}\label{equ:Riemannian distance and energy}
    \begin{aligned}
    d(x,x^*) = d(c) :=\int_0^1 \sqrt{\delta^\top_{c(s)}M(c(s))\delta_{c(s)}}ds,
    \end{aligned}
\end{equation}
where $\delta_{c(s)} := \frac{\partial c(s)}{\partial s}$.
The shortest path in Riemannian space, or \textit{geodesic}, between $x$ and $x^*$ is defined as 
\begin{equation}\label{equ:geodesic}
    \gamma(s) :=\argmin_{c(s)} {d(x,x^*)}.
\end{equation}

Leveraging Riemannian tools, one feasible feedback tracking controller for  \eqref{equ:pre cer sys}, can be obtained by integrating the differential feedback law \eqref{equ:pre dif fed} along the geodesic, $\gamma(s)$ \eqref{equ:geodesic}, as
\begin{equation}\label{equ:control integral}
    u_k = u^*_k + \int_0^1K(\gamma(s))\frac{\partial \gamma(s)}{\partial s}\,ds.
\end{equation}
Note, that this particular formulation is reference-independent, since the target trajectory variations do not require structural redesign of the feedback controller and is naturally befitting to the flexible manufacturing paradigm. Moreover, the discrete-time control input, $u_k$~\eqref{equ:control integral}, is a function with arguments ($x_k$, $x^*_k$, $u^*_k$) and hence the control action is computed from  the current state and target trajectory.

Then, under Definition~\ref{thm:pre ctr con} for a contracting system, e.g., \eqref{equ:pre cer sys} driven by \eqref{equ:control integral}, we have
\begin{equation}\label{inequ:distance bound}
        d(\gamma_{k+1}) \leq (1-\beta)^{\frac{1}{2}} d(\gamma_k),
\end{equation}
which states that the Riemannian distance between the state trajectory, $x$, and a desired trajectory, $x^*$ is shrinking (i.e., convergence of the state to the reference), with respect to the DCCM, $M$, with at least constant rate.

\subsection{System Description}
To address parametric uncertainties, for the remainder of this article, we consider the following discrete-time control affine nonlinear system with parametric uncertainty 
\begin{equation}\label{equ:uncertain control affine}
    x_{k+1} = f(r,x_k) + g(r,x_k)u_k,
\end{equation}
where vector $r$ represents the bounded uncertain parameters,
\begin{equation}\label{equ:parameter bound}
    r \in \mathcal{R} = \{r \in \mathbb{R}^\ell ~|~ r_{i,min} \leq r \leq r_{i,max} ~|~ i=1,\dots,\ell\},
\end{equation}
and functions $f$ and $g$ are smooth along the $x$ direction and Lipschitz continuous along $r$. The corresponding differential dynamics (and hence the contraction condition) can be determined for any specific value of the parameter $r$, i.e.,
\begin{equation}\label{equ:uncertain differntial system}
    \delta_{x_{k+1}} = A(r,x_k)\delta_{x_k} + B(r,x_k)\delta_{u_k},
\end{equation}
where $A(r,x_k):=\frac{\partial (f(r,x_k) + g(r,x_k)u_k)}{\partial x_k}$ and $B(r,x_k):=\frac{\partial (f(r,x_k) + g(r,x_k)u_k)}{\partial u_k}$. 
Hence, from Section \ref{sec:discrete-time contraction analysis}, a function pair ($M,K$), satisfying the contraction condition \eqref{equ:pre ctr con} for any $r \in \mathcal{R}$, i.e., satisfying
\begin{equation}\label{inequ:uncertain contraction condition}
\begin{aligned}
    \left(A_k(r)+B_k(r)K_k\right)^\top M_{k+1}\left(A_k(r)+B_k(r)K_k\right) \\
    - (1-\beta)M_{k} < 0, \qquad \forall r \in \mathcal{R}.
\end{aligned}
\end{equation}
ensures contraction of the uncertain system \eqref{equ:uncertain control affine} for the full range of uncertainty. 

\subsection{Objective and Approach}\label{sec:obj app}
The main objective is to ensure offset-free tracking of (\textit{a priori} unknown) time-varying references for an uncertain nonlinear system \eqref{equ:uncertain control affine}. Time-varying references are generated using an estimate of the uncertain parameter $\hat{r}_k$, such that the reference sequence $(x^*,u^*)$ satisfies the following
\begin{equation}\label{equ:triplet generating sys}
    x_{k+1}^* = f(r^*,x_k^*) + g(r^*,x_k^*)u_k^*,
\end{equation}
where $r^*_k = \hat{r}_k$.
Suppose that we have the desired state trajectory $(x^*_k,x^*_{k+1})$. The corresponding target control input, $u_k$, can be obtained, given a system model and an estimated parameter value, $\hat{r}_k$, by solution to \eqref{equ:triplet generating sys}. These solutions, $(x^*_k,x^*_{k+1},u^*_k)$, are only feasible solutions for the actual system dynamics \eqref{equ:uncertain control affine} when the estimated parameter, $\hat{r}$, is equal to the physical system value, $r$. Consequently, generating control references subject to parameter modelling error will result in incorrect control targets and hence state tracking offsets (see, e.g., \cite{wei2021ifacadchem}).
Thus, our ensuing objective is to force the parameter estimate, $\hat{r}$, to approach the real value, $r$, online, whilst ensuring stability (to reference targets) for the full range of parameter variation.
    
    
To ensure stability/convergence to any (feasible) reference trajectories for the full range of parameter variation, a contraction theory-based structure is imposed during training of a neural network embedded controller offline. Instead of using the process model to generate process data for general neural network training, we propose to use the model to directly learn the crucial information for the contraction-based control design (satisfying \eqref{inequ:uncertain contraction condition}): the contraction metric (DCCM), which implies how the process nonlinearity affects the contraction behavior; and a differential feedback gain, from which the control action is computed.  
This trained DCCM neural network is then embedded in a contraction-based structure for (state-feedback) real-time control providing stability guarantees across the full range of parametric uncertainty (``exploitation''). This then facilitates online learning of uncertain system parameters within the control-loop (``safe exploration'' see, e.g., \cite{shin2019reinforcement,vcrepinvsek2013exploration} for further discussion).
    

By integrating the power of well-studied, model-based modern control methods with the inherent ability of neural networks to handle system uncertainties, the proposed approach provides: (1) a systematic and efficient approach to embedding neural networks in the closed-loop control scheme; (2) certificates for stabilizability to arbitrary (feasible) time-varying references without structural redesign of the controller, by imposing a contraction-based controller structure and conditions; and (3) online model correction through iterative learning of uncertain system parameters.
}

\section{Neural Network Approach to Contraction Analysis and Control of Uncertain Nonlinear Systems}\label{sec:DCCM}
\RMA{The following sections detail the proposed neural network embedded contraction-based control with online parameter learning approach 
as follows.} Firstly, the family of models \eqref{equ:uncertain control affine} is used to generate the state data and local values of the Jacobian  matrices for training (with both arbitrary and specific distributions permitted). Secondly, the data set is fed into a neural network to learn the function pair $(M,K)$ satisfying \eqref{inequ:uncertain contraction condition}, using a tailored loss function. Then, the controller is constructed using the function pair $(M_{NN},K_{NN})$ (both of which are represented by the DCCM neural network) by implementing  \eqref{equ:control integral}. \LAI{Finally, a neural network-based parameter learning module is incorporated into the control-loop, to provide online estimation of uncertain parameters, as required for correct reference generation and offset-free tracking.}

\subsection{Model-based Data Generation}
The first step in the proposed methodology is to generate data, $\mathcal{D}$, from a family of system models \eqref{equ:uncertain control affine}. \RMA{As discussed in the Introduction, to operate a chemical process using random operating conditions to generate process data to learn an accurate model is infeasible due to stability/safety concerns. The idea proposed in the following is to use a model with uncertain parameters (which characterizes the inherent uncertain nonlinear nature of modern processes) to generate data, which can be done safely offline for an explicit range of uncertainty in the system model. The contraction-based analysis is performed for the full range of system uncertainty to ensure the contraction-based controller to be robust. In this way, provided the actual system model behaves inside the family of models considered, efficient and stabilizing control combined with online parameter learning can be achieved.}

In order to impose the contraction conditions~\eqref{inequ:uncertain contraction condition} during training, which utilizes the generated data set, consideration as to which parameters must be included in $\mathcal{D}$ is required. The Jacobian matrices, $A_k(r,x_k)$ and $B_k(r,x_k)$, can be explicitly calculated from the system model \eqref{equ:uncertain control affine} for specific $r$ values (see \eqref{equ:uncertain differntial system}). If the distribution for the uncertain parameter is known, then $r$ can be generated as a random variable with such a distribution to produce a more realistic data set for the uncertain model. Calculation of $M_{k+1}$ requires the possible next-step states (i.e., given a specific state, $x_k \in\mathcal{X}$, generate all possible next-step states, $x_{k+1} \in \mathcal{X}$, for all possible inputs, $u_k \in \mathcal{U}$, using \eqref{equ:uncertain control affine}). Consequently, the Jacobian matrices $A_k(r,x_k),B_k(r,x_k)$ and the two-step trajectories, $x_k,x_{k+1}$, under specific $r$, are needed in the data set, $\mathcal{D}$. Algorithm \ref{algo:data generation} summarizes the data set generation procedure.

\begin{algorithm}
    \SetAlgoLined
    Initialize $r,s,x_k$ and $u_k$ with lower bound values\\
    \For{$r \in \mathcal{R}$}{
        \For{$x_k \in \mathcal{X}$}{
            \For{$u_k \in \mathcal{U}$}{
                Calculate $x_{k+1} = f(r,x_k) + g(r,x_k)u_k$.\\
                Compute $A_k,B_k$.\\
                Store $\{r,x_k,x_{k+1},A_k,B_k\}_i$ in data set $\mathcal{D}$.
            }
        }
    }
    \caption{Data Set Generation.}
    \label{algo:data generation}
\end{algorithm}
\begin{remark}
The data set generation process can be accelerated by paralleling Algorithm~\ref{algo:data generation}, i.e. to calculate $\{x_{k+1},A_k,B_k\}$ with each $\{r,x_k,u_k\} \in \mathcal{R} \times \mathcal{X} \times \mathcal{U}$ in parallel.
\end{remark}
\begin{remark}\label{rem:datagen}
     An ideal data set would include all possible two-step-trajectories, $x_k,x_{k+1}$, and Jacobians, $A_k$, $B_k$, under all possible combinations of $r\in\mathcal{R}$, $x_k\in\mathcal{X}$, $u_k\in\mathcal{U}$. Naturally, numerical implementation of Algorithm \ref{algo:data generation} requires discretization of these continuous sets (see, e.g.,~\eqref{equ:parameter bound}) using a sufficiently small step size (forming, e.g., a mesh of states). The mesh can be nonlinear, depending on the nonlinearity of the system dynamics and additionally chosen to be finer near reference trajectories, i.e., to provide better accuracy when close to the desired state and corresponding control input. The condition on the mesh size to ensure contraction will be discussed in Section~\ref{sec:design analysis}
\end{remark}
\begin{remark}
\RMA{Straightforward extensions can be made to Algorithm~\ref{algo:data generation} such that the data set, $\mathcal{D}$, is generated whilst additionally considering measurement noise. For example, the next step state values, $x_{k+1}$, could be generated using the small state perturbation $x_k+\eta$, where $\eta$ denotes a bounded measurement noise variable. Consequently, the learned DCCM will inherently be capable of handling measurement noise, although this would require additional extension of the system model and hence contraction analysis that follows (e.g., via adaptation of the results in \cite{pham2009contraction}). Additionally, through straightforward modifications, guaranteed bounded disturbance responses could be shaped from the disturbance input to the state or output  (e.g., via the differential dissipativity approach of \cite{WangBao17}). Both the measurement noise accommodation and disturbance rejection extensions are omitted from this article to avoid over-complicating the presentation.}
\end{remark}
\subsection{DCCM Synthesis from Data} \label{sec:DCCMsynth}

\begin{figure}
    \begin{center}
       \includegraphics[width=\linewidth]{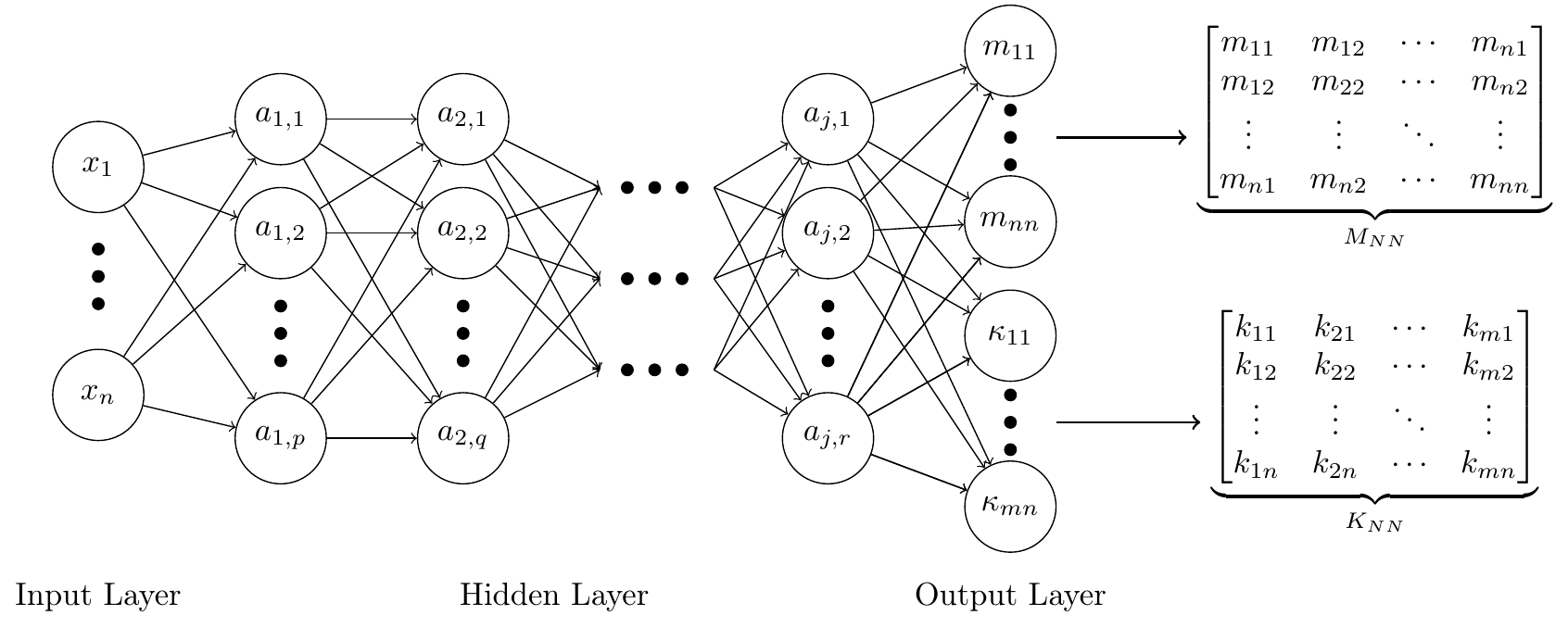}
        \caption{Illustration of the DCCM neural network structure.}
        \label{fig:GEN_NN}
    \end{center}
\end{figure}

In this work, a DCCM neural network is employed to represent the function pair ($M$,$K$) satisfying \eqref{inequ:uncertain contraction condition}. The structure of this DCCM neural network is shown in Fig. \ref{fig:GEN_NN}, whereby the inputs are the states of the system and the outputs are the numerical values of the matrices $M_{NN}$ and $K_{NN}$ for the corresponding states. Since the DCCM, $M_{NN}$, is a symmetric matrix, only the lower triangular components of $M_{NN}$ are of interest (i.e., only the lower triangular matrix is required to fully construct $M_{NN}$). Consequently, the first group of outputs are the components in the lower triangular matrix of $M_{NN}$ and the second group of outputs are the components of the controller gain $K_{NN}$ (see Fig. \ref{fig:GEN_NN}). Moreover, by exploiting the symmetric property of $M_{NN}$ the computational complexity is significantly reduced (only requiring $n(n+1)/2$ decision variables or network outputs). 

\subsubsection{Loss Function Design}

In order to train the DCCM neural network, a suitable loss function, $L$, is required. 
Inspired by the triplet loss in \cite{schroff2015facenet}, a novel (non-quadratic) objective loss function is developed herein to represent the positive definite properties of the neural represented metric function, $M_{NN}$, and contraction condition~\eqref{inequ:uncertain contraction condition}. By reforming \eqref{inequ:uncertain contraction condition}, we can rewrite the negative semi-definite uncertain contraction condition as a positive semi-definite condition (required for the subsequent loss function and training approach). Hence, we define $\Omega$ as (cf. \eqref{inequ:uncertain contraction condition})
\begin{equation}\label{equ:contraction condition omega}
   \Omega := -A_{cl,k}(r)^\top M_{{NN}_{k+1}}A_{cl,k}(r) + (1-\beta)M_{NN_k},
\end{equation}
where $A_{cl,k}(r):= A_k(r)+B_k(r)K_{NN_k}$. Then if $\Omega \geq 0$, the contraction condition \eqref{inequ:uncertain contraction condition} holds. Since these two conditions are inequalities of matrices, it is befitting to formulate the following loss function, $L$, based on quadratic penalty functions (see, e.g.,\cite{Bertsekas1976OnPenalty})
\begin{equation}\label{equ:LOSS_FUNCTION}
    \begin{aligned}
    L_{M_i} &=
    \begin{cases}
        -(|M_{NN_{(1,i)}}| - \epsilon_i)  &if ~(|M_{NN_{(1,i)}}| - \epsilon_i) \leq 0\\
        0 \ &else
    \end{cases}\\
    L_{\Omega j} &= 
    \begin{cases}
        -(|\Omega_{(1,j)}| - \epsilon_j) \quad \  &if ~(|\Omega_{(1,j)}| - \epsilon_j) \leq 0\\
        0 \ &else
    \end{cases}\\
    L &= \sum_i L_{Mi} + \sum_j L_{\Omega j},
    \end{aligned}
\end{equation}where $|M_{NN_{(1,i)}}|$ is the leading principle minor of $M_{NN}$ including the first $i$ rows and columns (i.e square submatrix) of matrix $M_{NN}$ and similarly for $\Omega_{(1,i)}$. Under Sylvester's criterion, by ensuring that each leading principle minor is positive, we can ensure the positive-definiteness of $M_{NN}$ and $\Omega$. A small positive value, $\epsilon_i$ or $\epsilon_j$, is introduced to reduce the effects of numerical errors, which may effectively return a semi-definite (or possible non-convergence) result. Each $L_{M_i}$ or $L_{M_j}$ returns a higher cost if the leading principle minor is  smaller than $\epsilon_i$ or $\epsilon_j$, otherwise, it returns zero. This encourages convergence of the leading principle minor to some value larger than $\epsilon_i$ or $\epsilon_j$. The loss function, $L$, (the sum of all $L_{Mi}$ and $L_{\Omega j}$), encourages all leading principle minors to be positive and hence the positive definiteness of both matrices, $M_{NN}$ and $\Omega$, which consequently implies $M_{NN}$ is a DCCM for the contraction of \eqref{equ:uncertain control affine}. Compared to existing CCM synthesis approaches using SoS programming, the proposed method permits contraction metric and feedback gain synthesis for non-polynomial system descriptions in addition to systems modeled with parametric uncertainty.

\subsubsection{DCCM Neural Network Training}

\begin{figure}
    \begin{center}
        \includegraphics[width=\linewidth]{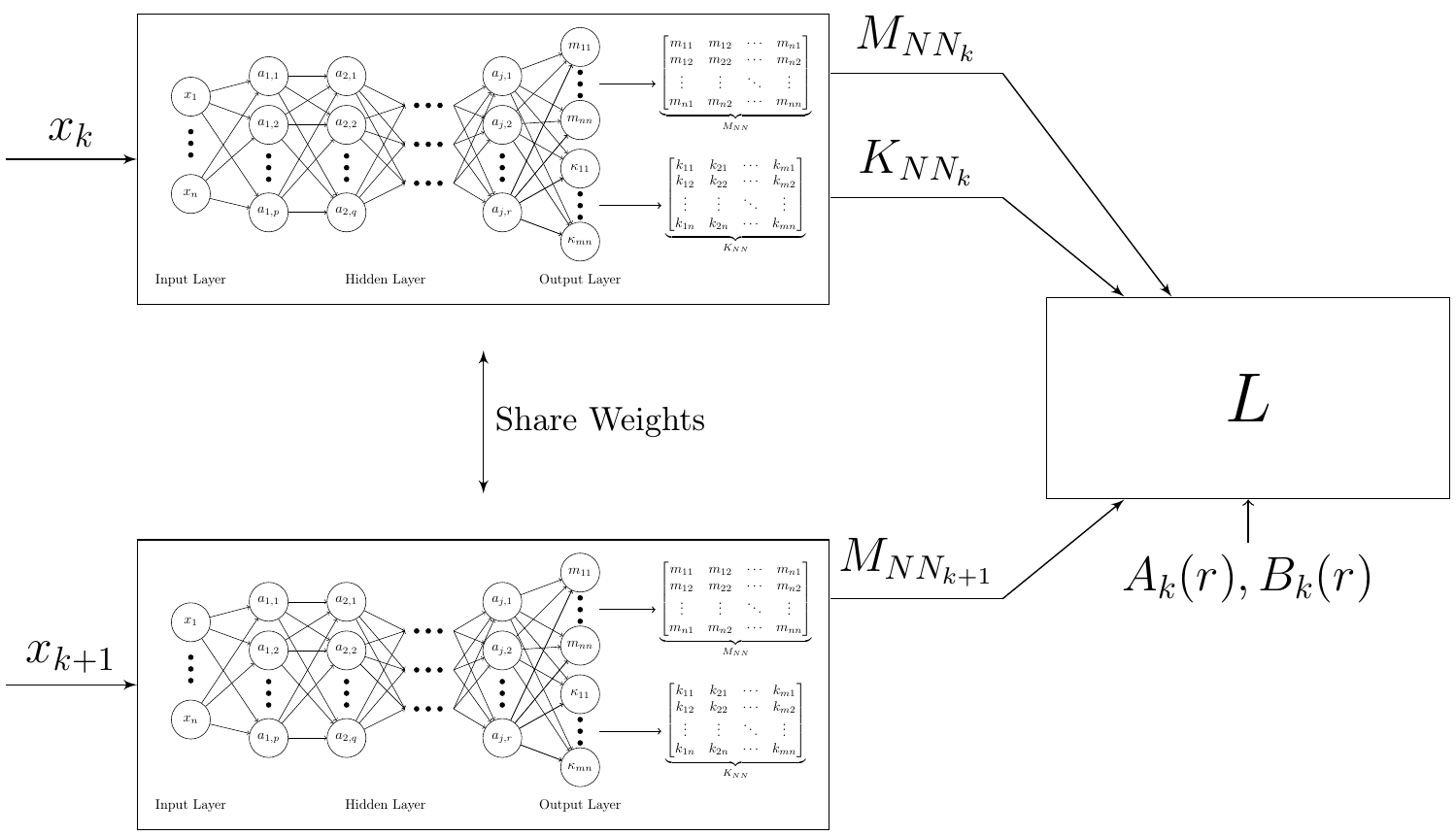}
        \caption{DCCM neural network training process block diagram.}
        \label{fig:siam}
    \end{center}
\end{figure}

Using the training data set, $\mathcal{D}$, generated from Algorithm \ref{algo:data generation} and loss function \eqref{equ:LOSS_FUNCTION}, we detail here the process for training the DCCM neural network function pair $(M_{NN},K_{NN})$. The DCCM neural network cannot be simply trained using the generalized structure in Fig.~\ref{fig:GEN_NN}, as the loss function \eqref{equ:LOSS_FUNCTION} requires both the DCCM neural network output under input, $x_k$, and also the output under input, $x_{k+1}$, since \eqref{equ:contraction condition omega} requires both $M_{NN_k}$ and $M_{NN_{k+1}}$ at the next step (i.e., $M_{NN}(x_{k+1})$ evaluated using $M_{NN}(x_k)$ and stepping forward using \eqref{equ:uncertain control affine}). To overcome this difficulty, we adopted a Siamese network (see Fig. \ref{fig:siam}) structure, whereby two neural networks, sharing the same weights, can be tuned at the same time, by considering both outputs of the weight sharing neural networks simultaneously. In addition, the Siamese network structure permits the use of outputs at time step $k$ and $k+1$ in the loss function. Furthermore, the learning can be done in parallel using a GPU to speed up the training, i.e. the complete set, $\mathcal{D}$ is treated as a batch, and a total loss, $L_t$, is defined by summing every loss function, $L_i$, where $L_i$ is the output from each element $\{r,x_k,x_{k+1},A_k,B_k\}_i$ in $\mathcal{D}$. Algorithm~\ref{algo:training} describes the training procedure for the Siamese neural network, with further details and discussion provided subsequently.

\begin{algorithm}
    \SetAlgoLined
    Stack elements, $\{r,x_k,x_{k+1},A_k,B_k\}_i \in \mathcal{D}$ as a batch\\
    \For{$iter \leq \ max \ iteration \ number$}{
        \For{each element $\{r,x_k,x_{k+1},A_k,B_k\}_i \in \mathcal{D}$}{
        Feed $x_k$ into the Siamese neural network.\\
        Feed $x_{k+1}$ into the Siamese neural network.\\
        Construct $M_{NN_k},M_{NN_{k+1}}$ and $K_{NN_k}$.\\
        Calculate the $i$-th element loss, $L_i$, as in \eqref{equ:LOSS_FUNCTION}.\\
        }
    Calculate total loss $L_t = \sum_i{L_i}$.\\
    Proceed backward propagation.\\ 
    \textbf{if} $L_t<\epsilon_{min}$ \textbf{then} Break.
    }
   
    Save $M_{NN}$ and $K_{NN}$.
    \caption{Training Procedure}\label{algo:training}
\end{algorithm}

The first step in Algorithm \ref{algo:training}, is to feed the two-step trajectories, $x_k,x_{k+1}$, into the Siamese networks, which have two sets of outputs, $M_{NN_k},K_{NN_k}$ and $M_{NN_{k+1}}$. Then, $M_{NN_k},K_{NN_k}$ and $M_{NN_{k+1}}$ are used to calculate the loss function, $L_i$ \eqref{equ:LOSS_FUNCTION} for  $\{r,x_k,x_{k+1},A_k,B_k\}_i$. 
The DCCM neural network is trained using backward propagation, whereby the loss is summed cumulatively at each iteration. As described in Algorithm~\ref{algo:training}, each iteration involves calculating the total loss, $L_t$, for the all elements in the data set; however, if the number of elements is sufficiently large, this process can be batch executed. The learning process is finally terminated when the total loss, $L_t$, is small enough, or the max number of predefined iterations is reached. The error threshold, $\epsilon_{min}$, is the smallest among all $\epsilon_i$ or $\epsilon_j$ in \eqref{equ:LOSS_FUNCTION}, which implies that provided the cumulative error is lower than this threshold, the contraction condition is satisfied for each point in the data set. 

\LAI{The computational complexity of the training process in Algorithm~\ref{algo:training} can be expressed in terms of the number of floating point operations (FLOPs). Feeding one element of the data set, e.g., the $i$-th element $\{r,x_k,x_{k+1},A_k,B_k\}_i$, through the Siamese network in Fig. \ref{fig:siam}, the computation requires $FLOPs = 2\sum_j{(2I_j-1)O_j} + 2N_{node} + 2n^3 + n^2 +2\sum_{i=1}^n FLOPs_{det_h}$, where $I_j$ and $O_j$ represent the number of inputs and outputs of layer $j$ respectively \cite{molchanov2019pruning}, $N_{node}$ represents the number of nodes of one neural network in Fig. \ref{fig:GEN_NN}, $n$ is the order of the system and $FLOPs_{det_h}$ represents the number of FLOPs for the determinant calculation of an $h\times h$  matrix (see, e.g.,~\cite{boyd2004convex} for more details).}

\begin{remark}
A number of existing strategies are available for quantifying the ``success" of a trained DCCM neural network, e.g., through testing and verification or statistical analysis and performance metrics. As an example, we note that the training methodology presented, is capable of incorporating direct validation by splitting or generating multiple data sets, say one for training, $\mathcal{D}_T$, and another for validation, $\mathcal{D}_V$, whereby each set can be given context specific weighting, pending, e.g., the target system trajectories or known regions of typical or safety critical operation. Due to the task specific nature of this process and range of techniques available, we have omitted its presentation here for clarity and refer the interested reader to \cite{taylor2006methods} for further details. Herein, and without loss of generality, we assume the training process was completed sufficiently, i.e., the training process was not stopped due to exceeding the maximum number of iterations, with a level of accuracy sufficient for the desired control application.  
\end{remark}

\subsection{Neural Network Embedded Contraction-based Controller}\label{sec:dccm to contr}
This section details the implementation of a contraction-based controller, of the form in~\eqref{equ:control integral}, that is obtained by embedding a neural network representation of the function pair $(M,K)$, i.e., $M_{NN}$ and $K_{NN}$ (calculated using Algorithms~\ref{algo:data generation} and~\ref{algo:training}). Foremost, the proposed neural network embedded contraction-based controller is described by (cf.~\eqref{equ:control integral})
\begin{equation}
\label{equ:control NN}
    u_k = u^*_k + \int_0^1K_{NN}(\gamma(s))\frac{\partial \gamma(s)}{\partial s}\,ds,
\end{equation}
where $(x_k^*,u_k^*)$ are the state and control  reference trajectories at time $k$. When the desired state value, $x_k^*$, changes, the feed-forward component, $u_k^*$, can be instantly updated, and the feedback component, $\int K_{NN} \delta_{\gamma}\,ds$, can be automatically updated through online geodesic calculation. Note that this approach results in setpoint-independent control synthesis. From \eqref{equ:Riemannian distance and energy} and \eqref{equ:geodesic}, the geodesic, $\gamma$, is calculated as
\begin{equation}\label{GEO_CAL}
\begin{aligned}
    \gamma(x,x^*) :&=\argmin_{c} {d(x,x^*)} \\&= \argmin_c \int_0^1{\frac{\partial c(s)}{\partial s}^T M_{NN}(c(s)) \frac{\partial c(s)}{\partial s}ds},
\end{aligned}
\end{equation}
where $M_{NN}$ and $K_{NN}$ are the function pair $(M,K)$ represented by the DCCM neural network (see Fig. \ref{fig:GEN_NN}), and recall from Section \ref{sec:discrete-time contraction analysis} that $c(s)$ is an $s$-parameterized smooth curve connecting $x$ ($s=0$) to $x^*$ ($s=1$). Implementing the contraction-based controller \eqref{equ:control integral} requires integrating the feedback law along the geodesic, $\gamma$, in \eqref{equ:geodesic}. 
Subsequently, one method to numerically approximate the geodesic is shown. 
Since  \eqref{GEO_CAL} is  an  infinite  dimensional  problem  over  all smooth curves, without explicit analytical solution, the problem must be discretized to be numerically solved. Note that the integral can be approximated by discrete summation provided the discrete steps are sufficiently small. As a result, the geodesic (\ref{GEO_CAL}) can be numerically calculated by solving the following optimization problem,
\begin{equation}\label{min:geodesic}
\begin{aligned}
    \bar \gamma(x,x^*) = \argmin_{\Delta \tilde{x}_{s}} &\sum_{i=1}^{N}{\Delta \tilde{x}_{s_i}^T M_{NN}(\tilde{x}_i) \Delta \tilde{x}_{s_i} \Delta s_i}\\
    s.t. \quad  & \tilde{x}_1 = x,~\tilde{x}_N = x^*,
\end{aligned}
\end{equation}where $\bar \gamma(x,x^*) \approx \gamma(x,x^*)$ represents the numerically approximated geodesic, $x$ and $x^*$ are the endpoints of the geodesic, $\tilde{x}_i$ represents $i$-th point on a discrete path in the state space, $\Delta \tilde{x}_{s_i} := \Delta {\tilde{x}_i} / \Delta {s_i} \approx {\partial c(s)}/{\partial s}$ can be interpreted as the displacement vector discretized with respect to the $s$ parameter, $\Delta \tilde{x}_s:=(\Delta \tilde{x}_{s_1},\dots,\Delta\tilde{x}_{s_N})$ is the discretized path joining $x$ to $x^*$ (i.e., discretization of c(s) in~\eqref{GEO_CAL}), all $\Delta {s_i}$ are small positive scalar values chosen such that $\sum_{i=1}^N \Delta s_i = 1$, $N$ is the chosen number of discretization steps (of s), $\tilde{x}_i = \sum_{j=1}^i \Delta \tilde{x}_{s_j} \Delta s_j + x$ represents the numerical state evaluation along the geodesic.
\begin{remark}
Note that \eqref{min:geodesic} is the discretization of \eqref{GEO_CAL} with $\Delta \tilde{x}_{s_i}$ and $\Delta_{s_i}$ as the discretizations of  $\frac{\partial c(s)}{\partial s}$ and $\delta_s$ respectively, whereby the constraints in \eqref{min:geodesic} ensure that the discretized path connecting the start, $x$, and end, $x^*$, state values align with the continuous integral from $s=0$ to $s=1$. Hence, as $\Delta_{s_i}$ approaches 0, i.e., for an infinitesimally small discretization step size, the approximated discrete summation in \eqref{min:geodesic} converges to the smooth integral in \eqref{GEO_CAL}.
\end{remark}
After the geodesic is numerically calculated using \eqref{min:geodesic}, the control law in \eqref{equ:control integral} can be analogously calculated using an equivalent discretization as follows
\begin{equation}\label{equ:controller}
u_k = u_k^* +  \sum_{i=1}^N \Delta \tilde{x}_{s_i} \Delta s_i K_{NN}(\tilde{x}_i).
\end{equation}

The state reference, $x^*$, is chosen to follow some desired trajectory, for which the corresponding instantaneous input, $u^*$, can be computed via real-time optimization methods (see ``Reference Generator'' in Fig.~\ref{fig:control block}), such that the triplet $(x_k^*,u_k^*,x_{k+1}^*)$, obtained from \eqref{equ:triplet generating sys}, satisfies \eqref{equ:uncertain control affine} 
for a specific value of the uncertain parameter, i.e., only when the modeled parameter matches the physical value, or $r^* = r$. The choice for this reference value, $r^*$, for the purpose of reference design, can be selected as the most likely or expected value for the uncertain parameter, i.e., $ r^* = \E[r]$. Hence, the corresponding desired control input at any time, $u^*_k$, can be calculated from \eqref{equ:triplet generating sys} as the expected corresponding control effort, $u^* =\E[u^*_k]$, given both the desired state values, $x^*_k,x^*_{k+1}$, and expected value for $r$. Suppose then, that there was some error (e.g., due to modeling) between the chosen uncertain parameter, $r^*$, and the exact value for $r$. Consequently, there will be some error when computing the corresponding control effort for the desired state trajectory (via \eqref{equ:uncertain control affine}), and moreover, for the resulting control effort in \eqref{equ:control NN}, denoted by $\tilde{u}_k = \bar{u}_k - u_k^*$, where $\bar{u}_k$ represents the control input reference generated using the correct parameter value $r$. The resulting \textit{disturbed} system, can be modeled, using \eqref{equ:uncertain control affine}, as
\begin{equation} \label{equ:deviation control affine}
    x_{k+1} = f(r,x_k) + g(r,x_k)(u_k + \tilde{u}_k),
\end{equation} 
where $u_k$ has the same form as \eqref{equ:control integral}. Inspired by the results in \cite{lohmiller1998contraction}, we have then have the following contraction result. 

\begin{lemma}\label{lemma:bnd}
    For the DCCM-based controller \eqref{equ:control integral} that ensures a system without uncertainty \eqref{equ:pre cer sys} is contracting, when parametric uncertainty is present \eqref{equ:deviation control affine}, the state trajectory, $x$, is driven by \eqref{equ:control integral} to the bounding ball around the target reference, $x^*$, as 
    \begin{equation}\label{eq:bball}
    d(\gamma_{k+1}) \leq (1-\beta)^{\frac{1}{2}} d(\gamma_k) + \sqrt{\alpha_2}G_k\|\tilde{u}_k\|.
    \end{equation}
\end{lemma}
\begin{proof}
A Riemannian space is a metric space, thus from the definition of a metric function (see e.g.,\cite{armstrong2013basic}) and from Definition \ref{thm:pre ctr con} we have the following inequality, 
\begin{equation}\label{equ:disturbance distance}
\begin{aligned}
    &d(\gamma(x_{k+1}, x_{k+1}^*)) = d(\gamma(\check{x} + g(r,x_k)\tilde{u}_k, x_{k+1}^*)) \\
                                & \leq d(\gamma(\check{x}_{k+1}, x_{k+1}^*)) + d(\gamma(\check{x}_{k+1} + g(r,x_k)\tilde{u}_k, \check{x}_{k+1})),
\end{aligned}
\end{equation}
where $\check{x}_{k+1} := f(r,x_k) + g(r,x_k)u_k$. Now, we consider the last two components of \eqref{equ:disturbance distance}. Firstly, from \eqref{inequ:distance bound}, we have $d(\gamma(\check{x}_{k+1}, x_{k+1}^*)) \leq (1-\beta)^{\frac{1}{2}}d(\gamma(x_{k}, x_{k}^*))$. Secondly, since the metric, $M_k$, is bounded by definition, then, $d(\gamma(\check{x}_{k+1} + g(r,x_k)\tilde{u}_k, \check{x}_{k+1})) \leq \sqrt{\alpha_2}G_k \|
\tilde{u}_k\|$, where $G_k = \max_{x_k} \| g(r,x_k) \|$. Thus we have the conclusion in \eqref{eq:bball}.
\end{proof}
\begin{remark}
The condition in \eqref{eq:bball} (cf.~\eqref{inequ:distance bound}) requires the disturbance term, $g_k \tilde{u}_k$ in \eqref{equ:deviation control affine}, to be bounded. The control deviation, $\tilde{u}_k$ is a constant, given a particular value for $r$, and it is a reasonable assumption that in control practice, the control-to-state mapping, $g_k$, is also bounded for all $k$.
\end{remark}
The choice for the uncertain parameter, $r^*$, when designing the reference trajectory, $(x^*,u^*)$, directly affects the radius of the ball to which the system \eqref{equ:pre cer sys} contracts, and naturally, a finite upper limit on the radius, due to this design choice, exists and can be described by the maximum disturbance, i.e., $\max \sqrt{\alpha_2}\|g_k \tilde u_k\|$. For continuity, note that by designing the reference about the expected value of the uncertain parameter, the expected radius of the bounding ball in \eqref{eq:bball} is zero and hence (cf. \eqref{inequ:distance bound}) recovers the undisturbed or exact contraction results of Section \ref{sec:discrete-time contraction analysis} (see specifically Definition~\ref{thm:pre ctr con}) when the expected parameter value correctly matches the physical value. Following this idea, we will present a method in the following section to adjust the reference parameter, $r^*$, such that it converges to the physical value, $r$, hence facilitating offset-free tracking.

\subsection{Online Parameter Learning}
\RMA{As Lemma \ref{lemma:bnd} provides conditions for closed-loop system stability for a range of uncertain parameters, it serves the foundation for ``safe exploration'' \cite{shin2019reinforcement,vcrepinvsek2013exploration}, through the addition of an online parameter estimation module, whilst maintaining stabilizing control. The additional parameter learning module (e.g., neural network training algorithm) is included in the closed-loop to learn the correct value of any uncertain parameters, such that correct reference generation and hence offset-free control can be obtained. Naturally befitting the existing approach, we present here a neural network based online parameter identification method. The estimation neural network is constructed as shown in Fig. \ref{fig:online nn}, whereby the input is the current system state, $x_k = (x_1,\cdots,x_n)^\top$, and the output is the uncertain parameter estimate, $\hat{r}_{k} = (\hat{r}_1,\cdots,\hat{r}_\ell)^\top$. This chosen neural network structure is a generalized treatment for parameter identification/estimation. It allows for extensions to state-dependent uncertain system parameters (e.g., the rate of a chemical reaction can be a function of the temperature (a state variable) in a chemical reactor) or a more general case that uncertain parameters can represent unknown/unmodeled system dynamics. 
\begin{figure}
    \begin{center}
        \includegraphics[width = \linewidth]{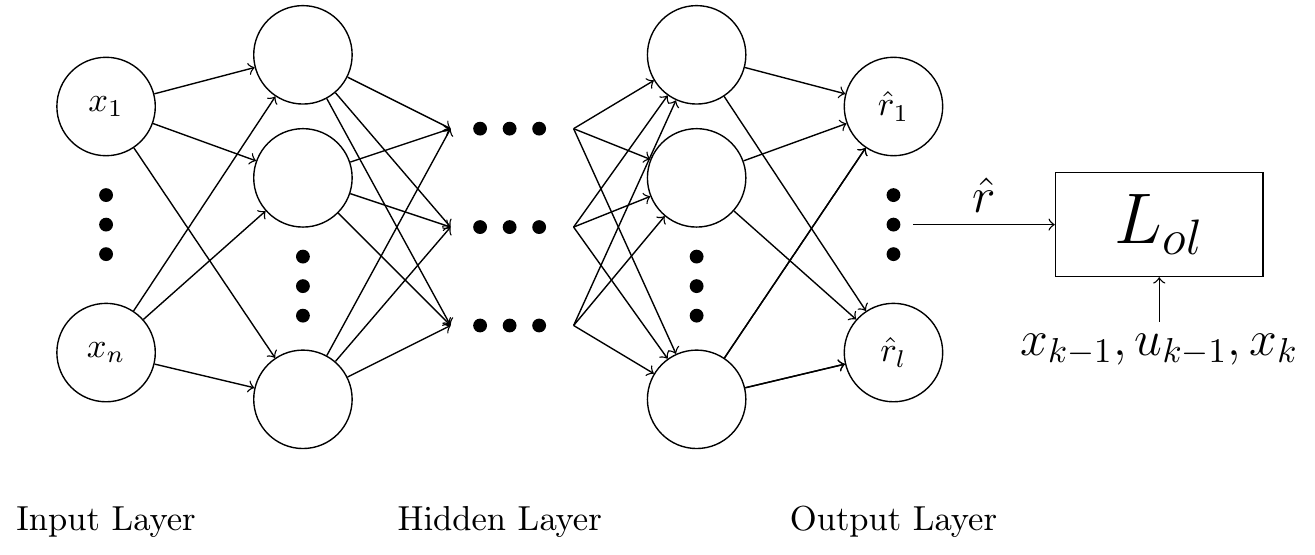}
        \caption{Illustration of the online parameter estimation neural network training process.}
        \label{fig:online nn}
    \end{center}
\end{figure}
To facilitate the learning process, as shown in Fig. \ref{fig:online nn}, a loss function, $L_{ol}$, is constructed to represent the error of prediction, defined as
\begin{equation}\label{equ:online loss}
    L_{ol} = \| x_k - f(\hat{r},x_{k-1}) - g(\hat{r},x_{k-1},u_{k-1}) \|.
\end{equation}
To clarify, $r^*$ is the value used for reference generation (as per Section~\ref{sec:dccm to contr}), which is updated by online parameter estimates, $\hat{r}$, for the physical system parameter, $r$, using the proposed Algorithm \ref{algo:online learning}. The initial estimate (used for reference generation), is taken as the expected (albeit potentially incorrect) value for the uncertain parameter, $\bar{r}$ (as per Section~\ref{sec:dccm to contr}).
\begin{algorithm}
    \SetAlgoLined
    Initialize the estimation neural network.\\
    \For{Each time step $k$}{
    Append $\{x_{k-1},u_{k-1},x_k\}_i$ to the training data-set, $\mathcal{D}_e$, for the estimation neural network.\\
    \For{$iter \leq max~iteration~number$}{
    \For{Each element $i$ in $\mathcal{D}_e$}{
        Calculate $L_{ol,i}$ using \eqref{equ:online loss}.\\
        Backpropagate using $L_{ol,i}$.\\
        }
    \textbf{if} $\max_i L_{ol,i} <\epsilon_{ol}$ \textbf{then} Break.
    }
    \uIf{$\hat{r}_k\in \mathcal{R}$ in \eqref{equ:parameter bound}}{
    Output $\hat{r}_k$.
    }
    \Else{
    Output $\hat{r}_{k-1}$.\\
    Reinitialize the estimation neural network.
    }
    }
    \caption{Online Parameter Learning}\label{algo:online learning}
\end{algorithm}
\begin{remark}
The learning algorithm needs sufficient non-repeating data (utilizing the reference model \eqref{equ:triplet generating sys} with the past state, control input, reference target and uncertain parameter values) to meet the minimum requirement for parameter convergence. Moreover, the amount of data required for identifiability increases with the dimensionality of the parametric uncertainty (see, e.g.,~\cite{olivier2017performance} for further discussion).
\end{remark}
Algorithm \ref{algo:online learning} describes the procedure of parameter estimation in the time interval $[k,k+1)$. Suppose $H$ denotes the number of available recorded time steps (historical data elements), then the data set, $\mathcal{D}_e$, contains the collection of elements $\{x_{k-H},x_{k-H+1},\cdots,x_{k-1}\}$. Since the parameter estimation neural network is trained using backward propagation, if the number of elements  is sufficiently large, this process can be batch executed (in parallel). The parameter learning process is finally terminated when the largest loss among elements in $\mathcal{D}_e$ is small enough (using the arbitrarily small threshold $\epsilon_{ol}$), or the max number of predefined iterations is reached, implying that the estimation error is sufficiently small for each historical element in the data set. Importantly, the estimated parameter, $\hat{r}_k$, is forced to lie inside the known bound $(r_{min},r_{max})$, satisfying \eqref{equ:parameter bound}, as required to ensure that the contraction-based controller maintains stability (as per Lemma \ref{lemma:bnd}). As the estimated parameter, $\hat{r}$, converges to the physical value, $r$, the reference model \eqref{equ:triplet generating sys} converges to that of the physical system \eqref{equ:uncertain control affine}, leading to the following conclusion for the closed-loop. 

\begin{corollary}\label{coro:offset free}
     The discrete-time nonlinear system \eqref{equ:uncertain control affine}, with neural network embedded controller \eqref{equ:control NN}, is stable with respect to a target reference (bounded convergence), in the sense of Lemma \ref{lemma:bnd}. Provided Algorithm \ref{algo:online learning} converges, the Riemannian distance between the system state and the desired reference additionally shrinks to zero.
\end{corollary}
\begin{proof}
The proof is straightforward by noting that the uncertainty results of Lemma \ref{lemma:bnd} collapse to that of Definition \ref{thm:pre ctr con} when the uncertain parameter $r$ is correctly identified. From convergence of Algorithm \ref{algo:online learning}, the reference generating model \eqref{equ:triplet generating sys} matches precisely the physical system \eqref{equ:uncertain control affine}, i.e., $r^*=\hat{r}=r$.  
\end{proof}

\subsection{Synthesis and Implementation Summary}
Lemma \ref{lemma:bnd} guarantees that a contraction-based controller (designed offline via Algorithms \ref{algo:data generation} and \ref{algo:training}) will at least drive an uncertain nonlinear system to a small neighborhood about any reference generated using the model in \eqref{equ:triplet generating sys}. Since stability is guaranteed (in the sense of boundedness about the target trajectory), we can update online the reference model using the identified parameter in Algorithm \ref{algo:online learning} (provided the estimation satisfies \eqref{equ:parameter bound}). As stated in Corollary \ref{coro:offset free}, when the reference model matches precisely the physical system, i.e., the parametric uncertainty is removed through online identification (estimation), the neural network embedded contraction-based controller \eqref{equ:control NN} also guarantees error free tracking. The proposed control synthesis and implementation (see Fig. \ref{fig:control block}) approach can be summarized as follows: 

\noindent\textbf{\textit{Offline:}}
\begin{enumerate}[label=\roman*.]
 \item Using the system model \eqref{equ:uncertain control affine}, generate training data, $\mathcal{D}$, via Algorithm \ref{algo:data generation}.
 \item Using the training data, $\mathcal{D}$, learn the metric, $M_{NN}$, and differential feedback gain, $K_{NN}$, via Algorithm \ref{algo:training}.
 \item Assign the reference model parameter, $r^*$, with the expected uncertain parameter value, $\bar{r}$, i.e., $r^* = \bar{r}$.
\end{enumerate}

\noindent\textbf{\textit{Online:}}
For each time step $k$
\begin{enumerate}[label=\roman*.]
 \item Update the identified parameters $\hat{r}_k$ using Algorithm \ref{algo:online learning} and generate the reference triplet, $(x_k^*,u_k^*,x_{k+1}^*)$ using the updated system model \eqref{equ:triplet generating sys} with $r^* = \hat{r}_k$. 
 \item Feed the state measurement, $x_k$, into the DCCM neural network to determine the current step metric and feedback gain, $M_{NN}$ and $K_{NN}$, as per Fig.~\ref{fig:GEN_NN}.
\item Calculate the numerical geodesic, $\bar{\gamma}(x_k,x_k^*)$, connecting the state, $x_k$, to the desired reference $x_k^*$, via solution to \eqref{min:geodesic} using the metric, $M_{NN}$.
 \item Using the geodesic information, $\bar{\gamma}_k$, differential feedback gain, $K_{NN}$, and the control reference, $u_k^*$, implement the control, $u_k$, via \eqref{equ:controller}.
\end{enumerate}

The proposed control approach is well suited for the dynamic control of modern industrial processes that are naturally highly nonlinear and modeled with uncertainty. Since the contraction metric, $M_{NN}$ (and thus the corresponding feedback control law \eqref{equ:controller}) is valid for the full range of parameter variation in $r$ (and hence $\hat{r})$, the parameter used by the reference generator, $r^*$, can be safely updated (explored) online simultaneously with control of the process. As a result, the proposed approach is capable of providing reference flexibility and efficient setpoint/trajectory tracking to ensure market competitiveness. Certificates for stability, to ensure safe and reliable operation, are explicitly detailed in the following section. The proposed contraction-based controller embeds neural networks, as shown in Fig. \ref{fig:control block}.} 
\begin{figure*}
    \begin{center}
        \includegraphics[width = 0.9\linewidth]{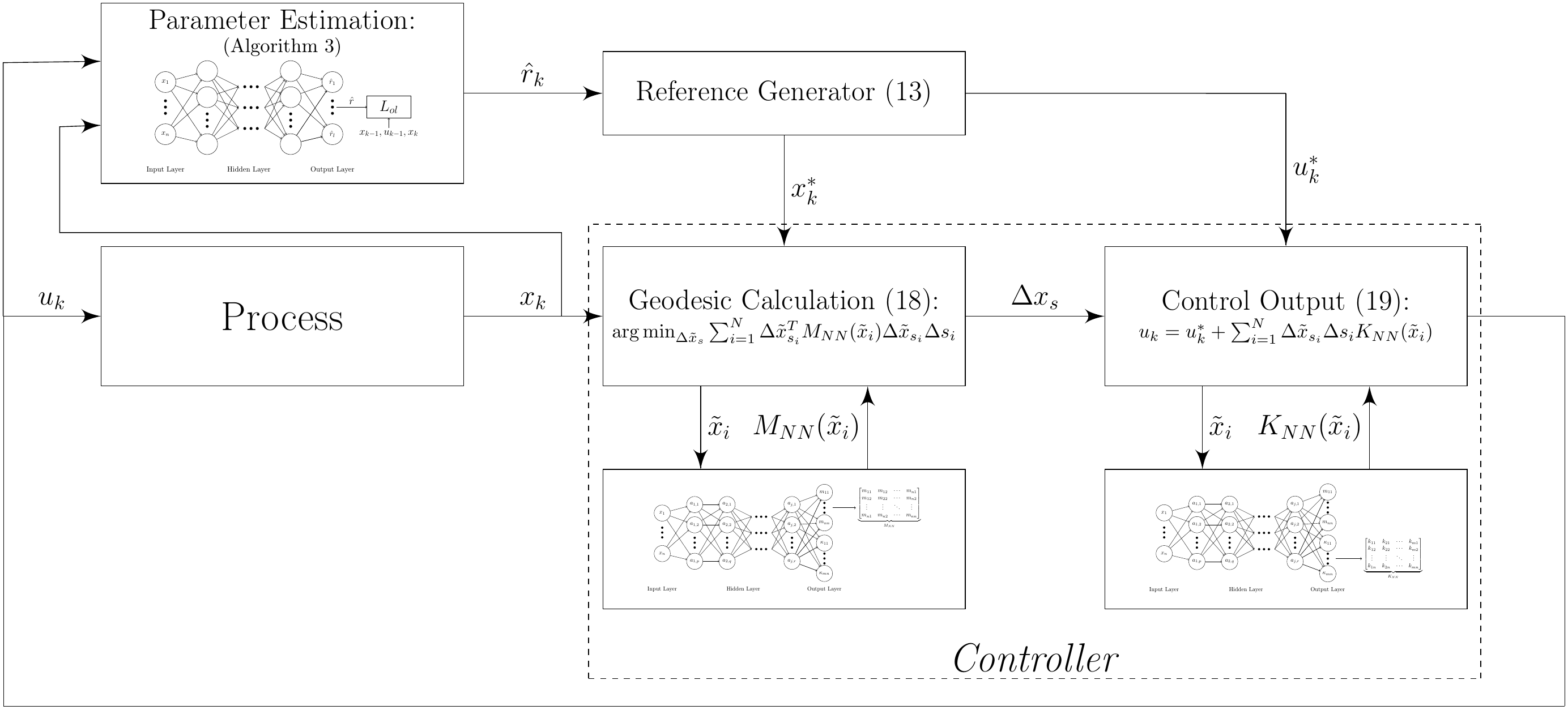}
        \caption{Proposed neural network embedded contraction-based control scheme.}
        \label{fig:control block}
    \end{center}
\end{figure*}

\section{Design Analysis}\label{sec:design analysis}
In practice, the DCCM neural network should be trained with a finite data set to make using Algorithms~\ref{algo:data generation} and \ref{algo:training} computationally tractable. The finite data set, $\mathcal{D}$, is comprised of a grid of points in $\mathcal{R} \times \mathcal{X} \times \mathcal{U}$, which naturally depends on the discretization step size or grid resolution (see Remark \ref{rem:datagen}). In this section, we develop bounding conditions on the contraction properties for the entire region of interest, when only a discrete number of data points are available. For clarity of presentation, we begin with considering the control affine system without uncertainty in \eqref{equ:pre cer sys}. The following theorem describes contraction regions in the form of a finite ball surrounding a known contracting point.

\begin{theorem}\label{thm:local contracting}
    If the contraction condition in \eqref{equ:pre ctr con} holds for a state and control value pair $(x^{\star},u^{\star})$ (satisfying \eqref{equ:pre cer sys}), then there exists a ball $B((x^{\star},u^{\star}),\xi)$ with radius $\xi=(\xi_x,\xi_u)$, centered at $(x^{\star},u^{\star})$,  in which the system \eqref{equ:pre cer sys} is \textit{locally} contracting with a rate no slower than $\lambda - L_{xu}\|\xi\|$, where $\lambda$ is the desired contraction rate and $L_{xu}$ is a Lipschitz constant.
\end{theorem}
\begin{proof}
Consider a function 
\begin{equation}\label{equ:max eig contraction}
\begin{aligned}
    h(x_k,u_k) := \max\eig ((\Theta_k^{-1})^\top A_{cl,k}^\top M_{k+1}A_{cl,k}\Theta_k^{-1} - I)),
\end{aligned}
\end{equation}
where $A_{cl,k} := A_k+B_kK_k$ and $M_k =: \Theta_k^\top \Theta_k$. Since all arguments of $h$ are assumed to be smooth, we can apply a Lipschitz condition to function $h$, yielding
\begin{equation}
  |h(x^{\star}+\xi_x,u^{\star}+\xi_u)-h(x^{\star},u^{\star})| \leq L_{xu}|\xi|,  
\end{equation} where $L_{xu}$ is a Lipschitz constant. By definition \cite{lohmiller1998contraction}, we have $h(x_k,u_k) \leq -\lambda$ . Hence, the largest variation of $h$ inside the ball can be upper bounded by
\begin{equation}
    h(x^{\star}+\xi_x,u^{\star}+\xi_u) \leq h(x^{\star},u^{\star})+L_{xu}|\xi|\leq -\lambda + L_{xu}|\xi|.
\end{equation}
Provided $-\lambda + L_{xu}|\xi| < 0$, the system \eqref{equ:pre cer sys} is contracting inside the ball, for which, there always exists a $|\xi|$ to ensure this negative condition. Moreover, the minimum contraction rate can be directly obtained by considering the maximum eigenvalue inside the ball.
\end{proof}

Theorem \ref{thm:local contracting} describes a local contraction property in the space $\mathcal{X} \times \mathcal{U}$. This property is generalized to the space $\mathcal{X}$ in the following extension, by considering all possible control values, $u \in \mathcal{U}$, for a particular state value, $x^\star \in \mathcal{X}$.
\begin{corollary}\label{coro:local contracting x}
     If the ball $B((x^{\star},u^j),\xi)$ centered at $(x^\star,u^j)$ forms a local contraction region for the system in \eqref{equ:pre cer sys} (for some control value $u^j$), and $B_x(x^{\star},\xi_x)\times\mathcal{U} \subseteq \underset{j}{\bigcup}B((x^{\star},u^j),\xi)$, then the system is locally contracting within $B_x(x^{\star},\xi_x)$ at $x^{\star}$.
\end{corollary}
\begin{proof}
From \eqref{equ:max eig contraction}, we have the contraction condition holds at different $u^j$ with radius $\xi_u$. If these balls are connected, then, there exists a ball around $x^{\star}$ such that $B_x(x^{\star},\xi_x)\times\mathcal{U} \subseteq \underset{j}{\bigcup}B((x^{\star},u^j),\xi)$.
\end{proof}

These results are extended in the following to systems with parametric uncertainties by considering locally contracting regions in the space $\mathcal{X} \times \mathcal{U} \times \mathcal{R}$ and hence the entire space of uncertainty, $\mathcal{R}$.

\begin{corollary}\label{coro:uncertain local contracting}
    If the contraction condition in \eqref{inequ:uncertain contraction condition} holds for the uncertain parameter value $r^{\star}$ with state and control pair $(x^{\star},u^{\star})$ for the system \eqref{equ:uncertain control affine}, then there exists a ball $B((x^{\star},u^{\star},r^\star),\xi)$ with radius $\xi=(\xi_x,\xi_u,\xi_r)$ centered at $(x^{\star},u^{\star},r^\star)$, for which the system \eqref{equ:uncertain control affine} is \textit{locally} contracting.
    Moreover, if $B_x(x^{\star},\xi_x)\times\mathcal{U}\times\mathcal{R} \subseteq \underset{j,\ell}{\bigcup}B((x^{\star},u^j,r^\ell),\xi)$, the system is locally contracting within $B_x(x^{\star},\xi_x)$ at $x^{\star}$.
\end{corollary}
\begin{proof}
These results are straightforward extensions of Theorem \ref{thm:local contracting} and Corollary \ref{coro:local contracting x}, by considering an additional parameter $r$ and hence dimension, i.e., $ \forall r \in \mathcal{R}$.
\end{proof}
By combining multiple locally contracting regions, a larger region of interest, $\mathcal{S}$, can be formed, for which we have the following immediate result.
\begin{corollary}\label{coro:union contraction region}
     If there exist multiple locally contracting regions $B_{x,i}:=B_x(x^i,\xi_{x,i})$ such that $S_x \subseteq \underset{i}{\bigcup} B_{x,i}$ (where $S_x \subseteq \mathcal{X}$ is an area of interest), then the area $S_x$ is a contraction region with the minimum contraction rate $\lambda_{\mathcal{S}_x,min}$ given by
    \begin{equation}
        \lambda_{\mathcal{S}_x,min} = \min_i (\lambda - L_{xur}||\xi_i||).
    \end{equation}
\end{corollary}
\begin{proof}
This result is straightforward from Theorem 2, by following a similar approach to the proof of Corollary \ref{coro:local contracting x}, where $h(x^{\star}+\xi_x,u^{\star}+\xi_u,r^{\star}+\xi_r) \leq -\lambda + L_{xur}|\xi|$ and $L_{xur}$ is a Lipschitz constant. The minimum contraction rate inside the region of interest is obtained by considering the maximum eigenvalue among local contraction regions covering the whole space of interest $S_x$.
\end{proof}

Theorem \ref{thm:local contracting} and Corollaries \ref{coro:local contracting x}--\ref{coro:union contraction region} state that the contraction property of a nonlinear system with parametric uncertainty can be determined by checking a finite number of local conditions (e.g., across a grid of state values).
In this way, a contraction rate close to the desired one can be achieved for an uncertain nonlinear system~\eqref{equ:pre cer sys} using finite data sets, hence making Algorithms \ref{algo:data generation} and \ref{algo:training} tractable.  
As the number of data points increases (and hence, considering increasingly small balls about each point), the minimum contraction rate for the unified region of interest, $S_x$, approaches the desired contraction rate.

\section{Illustrative Example}\label{sec:sim}

\begin{figure}
    \centering
    \includegraphics[width=\linewidth]{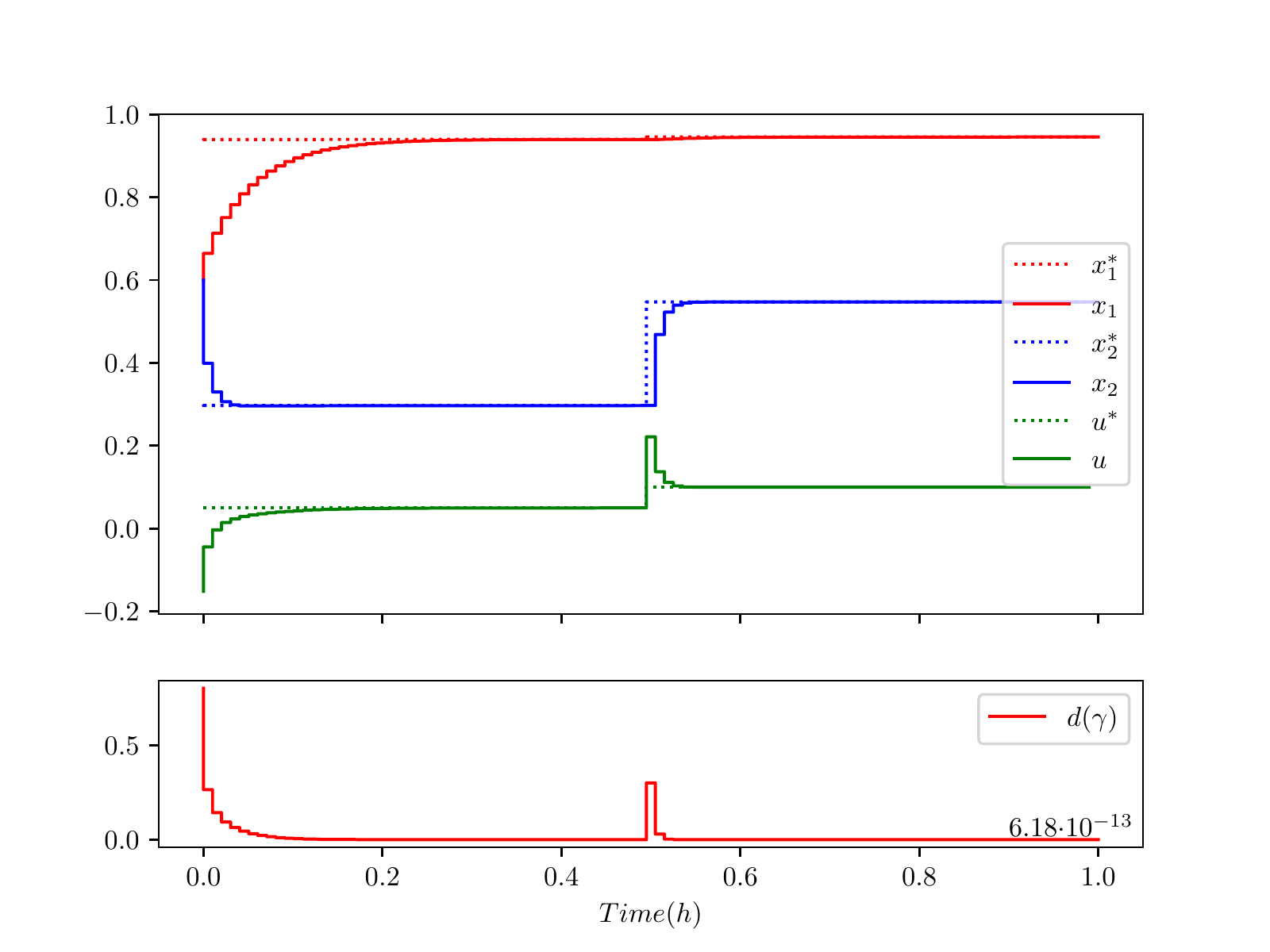}
    \caption{Simulation of CSTR control without parametric uncertainty.}
    \label{fig:cstr certain}
\end{figure}

\begin{figure}
    \centering
    \includegraphics[width=\linewidth]{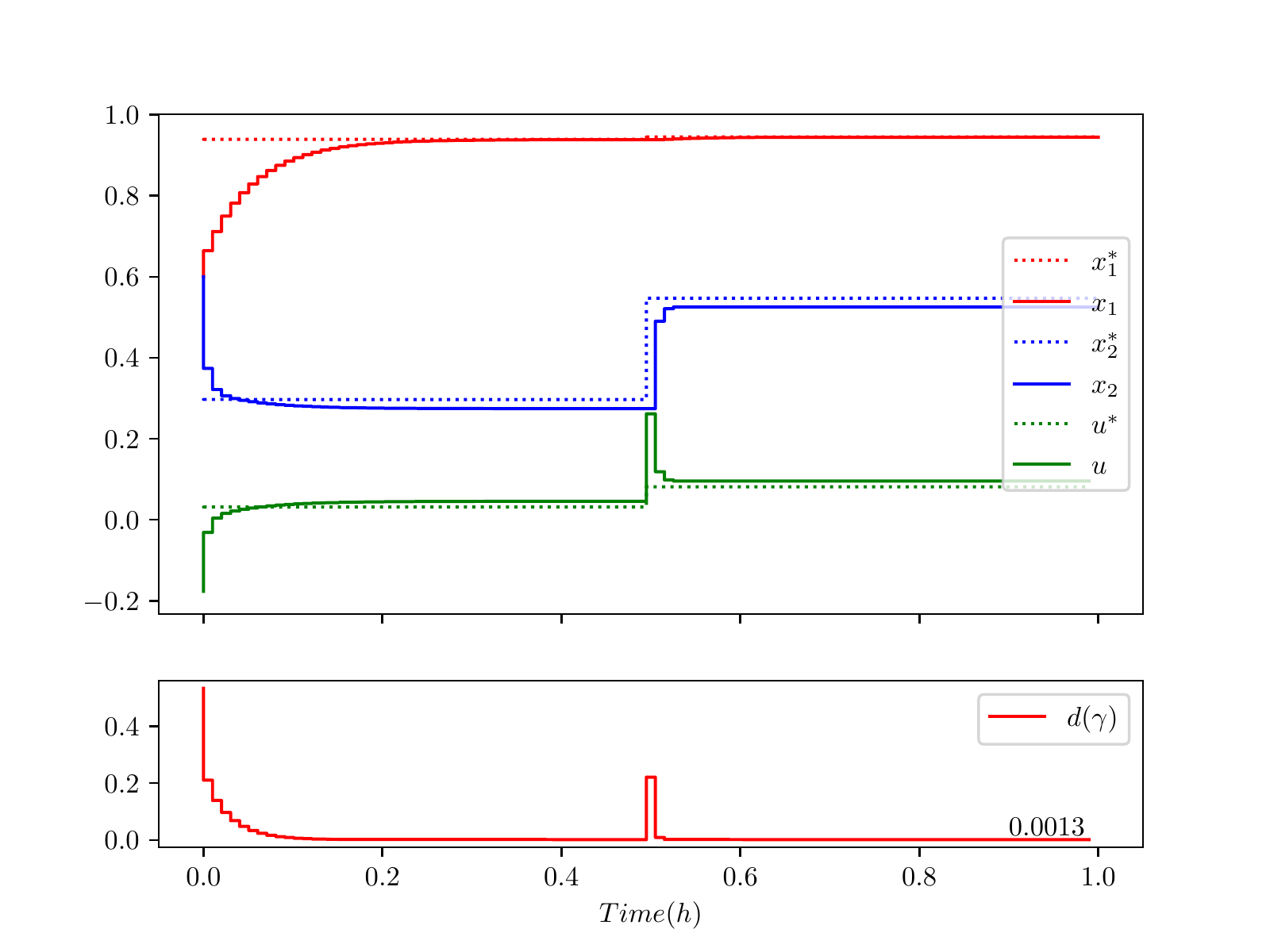}
    \caption{Simulation of CSTR control with parametric uncertainty and without parameter learning.}
    \label{fig:cstr uncertain}
\end{figure}

\begin{figure}
    \centering
    \includegraphics[width=\linewidth]{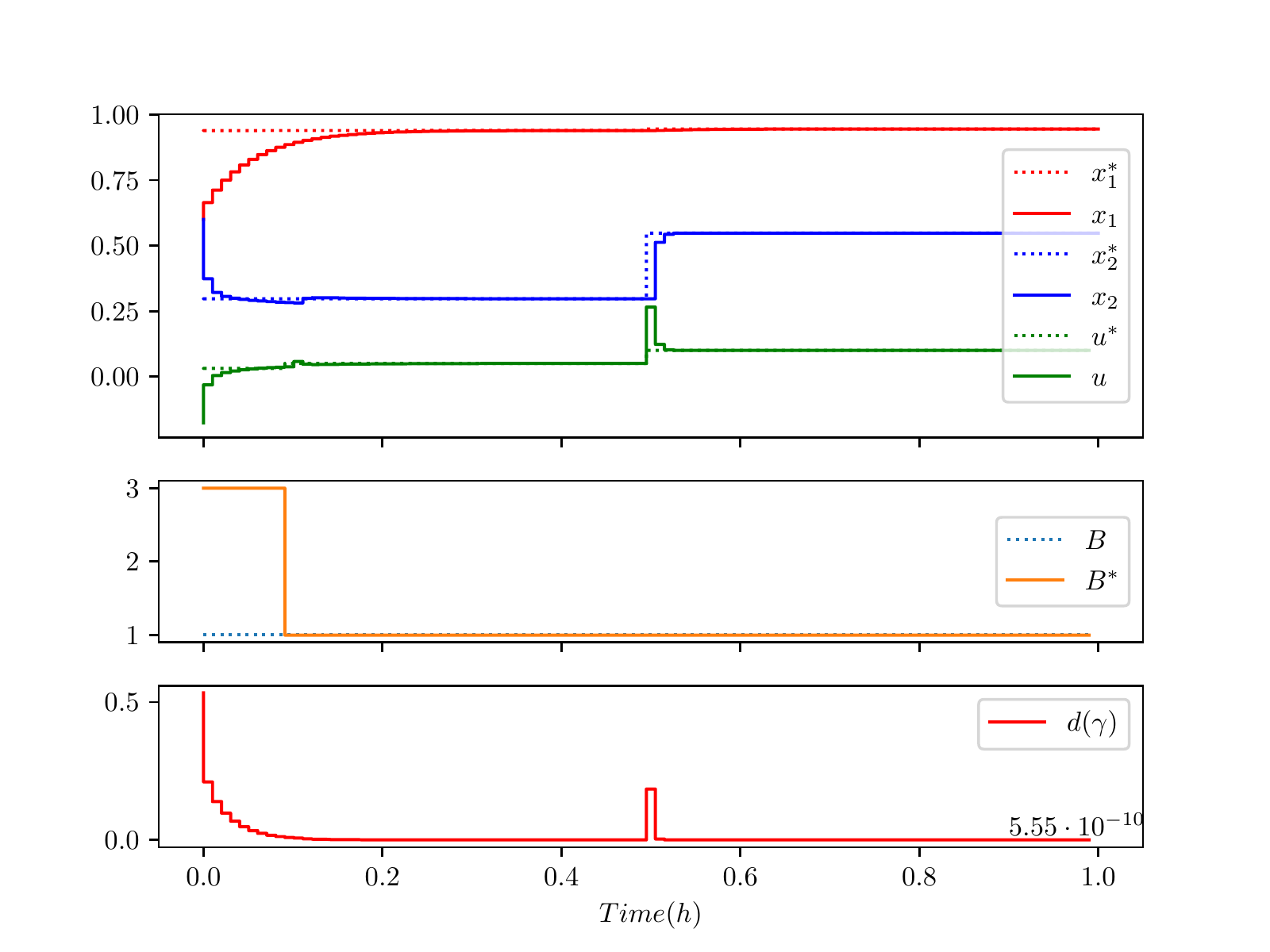}
    \caption{Simulation of CSTR control with parametric uncertainty and online parameter learning active from $t\geq0.1h$.}
    \label{fig:cstr uncertain learning}
\end{figure}

\RMA{To illustrate the proposed control design method and performance for both certain and uncertain nonlinear discrete-time systems, we present here two simulation examples which consider the following discrete-time model for a continuously stirred tank reactor (CSTR)  \cite{mccloy2021differential}:
\begin{equation}\label{equ:sim cstr certain}
         \begin{bmatrix}
         x_{1_{k+1}}\\ 
         x_{2_{k+1}}
         \end{bmatrix}= 
        \begin{bmatrix}
            \begin{aligned}
                &0.9x_{1_k} + 0.1 \phi_1(x_{1_k}) e^{\frac{\alpha x_{2_k}}{\alpha+x_{2_k}}} + 0.1(1-\zeta)x_{1_k}\\
                &0.9x_{2_k} + 0.1B \phi_2(x_{1_k}) e^{\frac{\alpha x_{2_k}}{\alpha+x_{2_k}}} + u_k
            \end{aligned}
        \end{bmatrix},
\end{equation}
where $\phi_i = Da_i(1-x_{1_k})$, $Da_1 = 1.25$, $Da_2 = 2.5$,  $\zeta = 0.1$, $\alpha = 0.8$ and the uncertain parameter $B \in [1,3]$ with the true value $B = 1$. The state and input constraints are $x_{1_k} \in [0.1,1.1]$, $x_{2_k} \in [0.1,1.1]$ and $u_k \in [-1,1]$, respectively. The normalized reactant concentration, reactor temperature and jacket temperature are denoted by $x_{1_k}$, $x_{2_k}$ and $u_k$, respectively. The time-varying state setpoints (based on the market demand and energy cost) are as follows:
\begin{equation}\label{eq:exstateref}
    (x_1^*(t),x_2^*(t)) = \begin{cases} (0.939,0.297), & \! \forall t\in[0,0.5) \\(0.945,0.547), & \! \forall t\in[0.5,1]\end{cases},
\end{equation}
whereby the control reference $u^*$ can be computed analytically using the system model \eqref{equ:sim cstr certain} and $B^*=B=1$ as $u^* = 0.050$~$\forall t\in[0,0.5)$ and $u^* = 0.1$~$\forall t\in[0.5,1]$. Similarly, when $B$ is incorrectly modeled, e.g., $B^*=3$, the (incorrect) control reference is computed as $u^* = 0.0312$~$\forall t\in[0,0.5)$ and $u^* = 0.0811$~$\forall t\in[0.5,1]$.  

Data generation was conducted offline using Algorithm~\ref{algo:data generation} via a square mesh of state ($x_k$), control ($u_k$), and uncertain parameter ($B$) values, with steps of $\frac{1}{60}$, $\frac{1}{10}$, and $\frac{1}{10}$ respectively. The DCCM and parameter estimation neural networks were both designed with ReLU hidden layer activation, linear input/output layer activation, a weight decay coefficient of $0.5$, and decay rates of $\beta_1=0.1$, $\beta_2=0.9$. The function pair $(M_{NN},K_{NN})$ was trained offline via Algorithm~\ref{algo:training} using the DCCM neural network structure as in Fig.~\ref{fig:GEN_NN} and Fig.~\ref{fig:siam} (with a learning rate of $0.05$, $3$ hidden layers, and $10$ neurons per hidden layer). For the system in this example, processing one element of data using the proposed neural network (shown in Fig.~\ref{fig:siam}) requires $3204$ FLOPs. To significantly reduce the training time, the DCCM neural network training algorithm was executed in parallel. Parameter estimates for $B$ were obtained via Algorithm \ref{algo:online learning} online, using the parameter estimation neural network as in Fig. \ref{fig:online nn} (with a learning rate of $0.00025$, $1$ hidden layer, and $4$ neurons per hidden layer). 


The CSTR was simulated using the proposed control design (as shown in Fig. \ref{fig:control block}) for the scenario when the reference generator uses the correct system parameter $B^*=B=1$ (hence, online learning is not required). Fig.~\ref{fig:cstr certain} shows that when the exact model is known the discrete-time neural network embedded contraction-based controller~\eqref{equ:controller} is capable of offset-free tracking. 
The system was then simulated with incorrect value of $B$ (with $B^*=3$), and without online parameter learning. Fig.~\ref{fig:cstr uncertain} shows that bounded tracking was achieved as per Lemma \ref{lemma:bnd} (observe that the Riemannian geodesic distance, $d(\gamma(x,x^*))$, converges to a non-zero value comparatively), whereby the incorrect control reference $u^*$ generated using the incorrect value for $B$ caused the tracking offsets (see Section \ref{sec:dccm to contr}). To further demonstrate the overall control approach, the same incorrectly modeled system was then simulated with the online parameter estimation module active from time $t \geq 0.1h$. As shown in Fig.~\ref{fig:cstr uncertain learning}, the proposed approach achieved bounded reference tracking (as per Lemma~\ref{lemma:bnd}) when parametric uncertainty was present (see $t \in [0,0.1h)$), and after the online parameter learning algorithm converged, offset free tracking (see also $d(\gamma(x,x^*)) \to 0$) was achieved as per Corollary \ref{coro:offset free}. 
}


\section{Conclusion}\label{sec:conclusion}
\JB{In this article, a framework was developed to train a DCCM neural network (contraction metric and feedback gain) for contraction analysis and control using a nonlinear system model with parametric uncertainties.} Considerations were made for the discrete-time contraction and stability for certain nonlinear systems, which for known bounds on modeling uncertainty, were then extended to provide direct analysis and controller synthesis tools for the contraction of uncertain nonlinear systems. \RMA{An online parameter ``safe'' learning module was also included into the control-loop to facilitate correct reference generation and consequently offset-free tracking.} The resulting contraction-based controller, which embeds the trained DCCM neural network, was shown capable of achieving efficient tracking of time-varying references, for the full range of model uncertainty, without the need for controller structure redesign. 
\ifCLASSOPTIONcaptionsoff
  \newpage
\fi
\bibliographystyle{IEEEtran}
\bibliography{NN-DCCM}
\end{document}